\documentclass[12pt]{article}

\usepackage[colorlinks,linkcolor=blue,citecolor=blue,urlcolor=blue,bookmarks,bookmarksnumbered]{hyperref}
\usepackage{amsmath,amssymb,amsfonts}
\usepackage[paper=letterpaper,margin=1.0in]{geometry}
\usepackage{graphicx,xcolor}
\usepackage{braket}
\usepackage{caption}
\usepackage{subcaption}
\usepackage{cite}
\usepackage{mathrsfs}

\newcommand{\be}{\begin{equation}}
\newcommand{\ee}{\end{equation}}
\newcommand{\bea}{\begin{eqnarray}}
\newcommand{\eea}{\end{eqnarray}}
\newcommand{\ba}{\begin{array}}
\newcommand{\ea}{\end{array}}
\newcommand{\ben}{\begin{enumerate}}
\newcommand{\een}{\end{enumerate}}
\newcommand{\bi}{\begin{itemize}}
\newcommand{\ei}{\end{itemize}}
\newcommand{\bc}{\begin{center}}
\newcommand{\ec}{\end{center}}
\newcommand{\bfig}{\begin{figure}}
\newcommand{\efig}{\end{figure}}
\newcommand{\bq}{\begin{quotation}}
\newcommand{\eq}{\end{quotation}}
\newcommand{\bt}{\begin{table}}
\newcommand{\et}{\end{table}}
\newcommand{\btab}{\begin{tabular}}
\newcommand{\etab}{\end{tabular}}
\newcommand{\bs}{\begin{slide}}
\newcommand{\es}{\end{slide}}

\newcommand{\IR}{\mathbb{R}}

\let\ba=\overline

\let\d=\delta

\def\rd{{\rm d}}

\let\g=\gamma

\let\j=\psi

\def\IR{\relax\leavevmode{\rm I\kern-.18em R}}
\def\ZZ{\relax\leavevmode
       \ifmmode\mathchoice
       {\hbox{\sf Z\kern-.4em Z}}
       {\hbox{\sf Z\kern-.4em Z}}
       {\lower.9pt\hbox{\scriptsize\sf Z\kern-.36em Z}}
       {\lower1.2pt\hbox{\tiny\sf Z\kern-.36em Z}}
       \else{\sf Z\kern-.4em Z}\fi}
\def\RR{\relax\leavevmode
       \ifmmode\mathchoice
       {\hbox{\sf R\kern-.4em R}}
       {\hbox{\sf R\kern-.4em R}}
       {\lower.9pt\hbox{\scriptsize\sf R\kern-.36em R}}
       {\lower1.2pt\hbox{\tiny\sf R\kern-.36em R}}
       \else{\sf R\kern-.4em R}\fi}

\def\resetby#1#2{\@addtoreset{#2}{#1}}
\def\seceq{\@addtoreset{equation}{section}
              \def\theequation{\thesection.\arabic{equation}}}

\def\Label#1{\label{#1}%
                \smash{\hbox to0pt{\raise1ex\hbox{\tiny[#1]}\hss}}}
\def\noLabels{\let\Label=\label}

\makeatletter
\DeclareRobustCommand\widecheck[1]{{\mathpalette\@widecheck{#1}}}
\def\@widecheck#1#2{%
    \setbox\z@\hbox{\m@th$#1#2$}%
    \setbox\tw@\hbox{\m@th$#1%
       \widehat{%
          \vrule\@width\z@\@height\ht\z@
          \vrule\@height\z@\@width\wd\z@}$}%
    \dp\tw@-\ht\z@
    \@tempdima\ht\z@ \advance\@tempdima2\ht\tw@ \divide\@tempdima\thr@@
    \setbox\tw@\hbox{%
       \raise\@tempdima\hbox{\scalebox{1}[-1]{\lower\@tempdima\box\tw@}}}%
    {\ooalign{\box\tw@ \cr \box\z@}}}
\makeatother

 \allowdisplaybreaks

\begin{document}

{\footnotesize
${}$
}

\begin{center}


{\Large \bf Physical limits on information metrics and quantum gravity as gravitized quantum theory}\\


\renewcommand{\thefootnote}{\fnsymbol{footnote}}

{\bf Per~Berglund}${}^{1}$%
 \footnote{\raggedright\baselineskip=13pt per.berglund@unh.edu,~~
            \footnotemark andrew.geraci@northwestern.edu,~~
            \footnotemark thubsch@howard.edu, 
            \hglue1.1pc 
            \footnotemark david.mattingly@unh.edu,~~
            \footnotemark dminic@vt.edu (Corresponding author)
            \addtocounter{footnote}{-4}},
{\bf Andrew~Geraci}${}^{2}$\footnotemark, 
{\bf Tristan~H{\"u}bsch}${}^{3}$\footnotemark,
{\bf David~Mattingly}${}^{1}$\footnotemark,
and~{\bf Djordje~Minic}${}^{4}$\footnotemark



\begin{small}
\baselineskip=15pt\it
${}^1$Department of Physics and Astronomy, University of New Hampshire, Durham, NH\\
${}^2$Department of Physics and Astronomy, Northwestern University, Evanston, IL\\
${}^3$Department of Physics and Astronomy, Howard University, Washington, DC\\
${}^4$Department  of Physics, Virginia Tech, Blacksburg, VA\\

\end{small}

\end{center}


\begin{abstract}
There is a long history in both general relativity and quantum mechanics of removing fixed background structures, thereby making observed objects and measurement processes dynamical.  We continue this evolution 
by combining central insights from both theories 
to argue that physical limits on information collection resulting from quantum gravity coupled with general covariance preclude the fixed information geometry still assumed in both information theory and quantum mechanics.  As a consequence there must be a gravitized, generally covariant extension of both theories. We also propose a novel experimental test involving intrinsic triple and higher order quantum interferences that would provide evidence for dynamical information metrics and a dynamical Born rule.
\end{abstract}

\begin{center}
Essay written for the Gravity Research Foundation
2025 Awards \\ for Essays on Gravitation.\\
Submission date: March 29, 2025.
\end{center}

\newpage

As 
physicists struggle to construct a conceptually sound, mathematically rigorous, and experimentally tested theory of quantum gravity~\cite{deBoer:2022zka}, 
the differences between quantum mechanics and general relativity are often accentuated:
quantum mechanics is a theory of the small while general relativity controls the universe writ large; 
relativity is a theory of determinate smooth geometry, whereas quantum mechanics is a theory of operators and measurement indeterminacy.  
Such framings, while perhaps helpful to 
categorize the theoretical challenges and  
assuage our wounded egos after a century of only ever partial successes,
%
quite definitely obscure the similarities between the theories. 
In particular, both theories contain conceptually deep but \textit{incompletely realized} revolutions about how science should separate the observed 
dynamical systems 
for which we wish to establish physical laws, 
from the observer ---
the outside static entity who defines the underlying structure 
within which observed dynamics happens.
Both paradigms move towards minimizing any separation: 
after all, ideally, a physical theory should {\em dynamically\/} define 
both the observer and the observed --- as well as the demarcation between them.
Alas, neither paradigm achieves this goal fully. 

In this essay we take a step towards completing this dynamic revolution: 
We combine insights from both theories to information theory, which implicitly contains idealized non-dynamical operations and observers, show how these insights suggest a gravitized and generally covariant extension of quantum mechanics, and propose a novel experimental test.  This moves both theoretical quantum gravity~\cite{deBoer:2022zka} and quantum gravity phenomenology~\cite{Addazi:2021xuf} forward in significant, qualitatively new ways.

The story of dynamical versus absolute objects is well-known in both general relativity and quantum mechanics, although the incompleteness and links between the two are perhaps under-appreciated.  Einstein, influenced by Mach's ideas about spacetime being totally dynamically determined by the distribution of matter, established general covariance and dynamism as a defining principle of general relativity. His goal was to remove the ``fixed stage,''  the absolute, ethereal, immutable spacetime framework of Newtonian physics within which dynamical geometry happens.  Indeed, Einstein considered this removal the deepest insight of relativity. In the end, however, general relativity only turned out to contain an \textit{incomplete} implementation of Machian principles, as  maximally symmetric de Sitter/anti-de Sitter spacetimes exist without matter, counter to  Einstein's original hopes.

Quantum mechanics also explicitly adjusts the relationship between observed and observer 
presupposed in classical Newtonian mechanics. Bystanding observers who can witness a dynamical system and yet leave it undisturbed and unchanged
were 
understood to be figments of wishful thinking.
Measurement itself was folded into the system dynamics, the act of measurement can change a quantum state, yet the current paradigm is still a very incomplete dynamical revolution. Quantum mechanics assigns states to classical probabilities post measurement via the Born rule, which is an independent 
axiom 
of the theory (analogous to the the famous fifth postulate of Euclidean geometry, which was generalized in the context of non-Euclidean geometries).  
In geometric quantum mechanics, the Born rule is encapsulated by the maximally symmetric metric on complex projective spaces, the Fubini--Study metric~\cite{Ashtekar:1997ud}. It is a fixed, non-dynamical metric in Hilbert space, clearly in direct contradiction with the spirit of general covariance and the absence of fixed geometric backgrounds.  Ironically, the Fubini--Study metric solves the Einstein equation in complex projective space with a cosmological constant like term, which means that in both quantum mechanics and general relativity the obstruction to full dynamism is geometrically similar.

The Fubini--Study metric is not, however, fundamentally a contradiction with general covariance that comes from quantization --- it is not a quantum anomaly.  Rather, it is 
inherited 
from classical probability theory and classical information geometry.
The Born rule is strongly constrained by requiring post-measurement probabilities to follow the usual rules of inference. In classical frequentist probability theory, the distance between distributions is governed by the maximally symmetric Fisher metric, which is 
by \v{C}encov's theorem the unique information metric {\em under sufficient statistics and identical independent measurements}~\cite{Cencov:1981Sta,
Fujiwara:2024Hom}. A sufficient statistic is a set of parameters derived from data that allow one to reproduce the probability distribution, e.g. the mean and standard deviation constitute a minimal sufficient statistic for a Gaussian distribution. The full data set is also always a sufficient statistic.  As should be clear from the above, \v{C}encov's theorem relies on the assumptions, typical in probability analysis, that:
({\small\bf1})~data is a set of permanent results from independent identically distributed (i.i.d.) measurements, and 
({\small\bf2})~sufficient statistics exist. Both these assumptions fail in quantum gravity in very specific ways. {\it As a result, the question of whether the Born rule is fixed or dynamical in quantum gravity becomes an empirical question.} 
And somewhat shockingly, tests of the Born rule are very weak with constraints on deviations only one part in $10^3$~\cite{Berglund:2023vrm}.

Let us consider the first assumption, which is one we make in science almost without thought~\cite{Berglund:2025qyo}.   For an information metric to be defined everywhere over a probability space, one needs to be able to run an experiment many times.  If each measurement is i.i.d. and the probability distributions have a finite sufficient statistic this poses little conceptual problem: the experimenter simply updates the statistic with each experimental run, stores the statistic in a permanent manner, and zeroes in on the ``correct'' distribution.  If the measurements or data storage back-react however, the construction falls apart, and this happens in gravity.  Let $X$ constitute permanent physical data about some system $S$. $X$ therefore must be stored in some device $D$ that incorporates a distinguishable distribution of conserved charge. With energy eigenstates or spin, the state of $D$ for each measurement $x \in X$ will gravitate and by the equivalence principle back-react on $S$ on each experimental run in a distinguishable way.  Therefore $X$ will not be i.i.d., but instead non-Markovian~\cite{Berglund:2025qyo}, an effect closely related to the ``memory loophole'' in quantum foundations and violations of the CHSH inequality~\cite{
Barrett:2002Qua}. 

One could try to evade this by employing gauge charges which don't directly couple to gravity. A global charge has no associated local gauge field, and hence $D$ would never affect $S$, allowing successive observations to remain independent.  However in \textit{quantum gravity} there are no global charges, only dynamical charged objects coupled to gauge fields~\cite{Harlow:2018tng, Heckman:2024oot}. The non-zero energy-momentum tensor from the gauge fields due to the charged source will back-react on $S$ on each experimental run. One cannot either simply turn down the magnitude of the charge in $D$ arbitrarily to minimize the amplitude of the gauge field and its energy momentum tensor, as such a process would violate the weak gravity conjecture~\cite{Harlow:2022ich}.   Therefore in quantum gravity the data in $X$ is not i.i.d. but instead non-Markovian~\cite{Berglund:2025qyo}, as each successive recording of a data point affects the next measurement.

We now turn to the second assumption, that sufficient statistics exist~\cite{Berglund:2025qyo}.  For a single observer this is certainly true, but the situation is more subtle with multiple observers.  Science usually assumes the following scenario: Consider two observers Alice and Bob, each with an identical copy of a local system $S$. $S$ is theoretically governed by a physical model parameterized by some set of parameters $\theta_i \in \Theta$. Alice takes a series of measurements $D$ on her copy.  Given enough data points $d_i \in D$ to make a sufficient statistic, Alice can compare with the theoretical predictions for the predicted distribution of the values of $d_i$ and infer the corresponding values of the $\theta_i$.  This is what we do when we e.g.\ run a regression.  The amount of information Alice gains about $\Theta$ with each $d_i$ is captured by the Fisher information.  Now, consider Alice sending the information in $D$ to Bob that allows him to build an equivalent statistic $D'$. We assume Bob can use $D'$ to make the same inferences about $\Theta$, the information is invariant under mapping from one sufficient statistic to another. This invariance dictates that Bob's information metric is also the Fisher metric. Additionally, Bob's inferences should match what he would infer from measurements of his copy of $S$.  This principle, which we have dubbed ``local information objectivity''~\cite{Berglund:2025qyo}, is a core tenet of science --- we spend vast amounts of time and money running multiple independent experiments and comparing data and theoretical inferences as part of our standard scientific method. The corresponding invariance of information inference for all observers implies that there is no dynamics possible for the Fisher metric.

The flaw in the above argument is that {\it in a generally covariant gravitational theory there is no such thing as purely local data} --- gauge invariant operators are gravitationally dressed, naturally extend out to infinity, and hence are not unique, but depend on unmeasured boundary conditions in a way that cannot be removed~\cite{Berglund:2025qyo}.  Therefore, in gravity, the complete set of ``data'' necessary to prove uniqueness of the Fisher information metric constitutes not only local recorded experimental data $D$, but also a set of \textit{assumed} boundary conditions $C$.   

In a non-gravitational theory dependence on \textit{unknown} boundary conditions can be minimized or removed altogether. Vanishing boundary conditions can be physically imposed --- we can place an electromagnetic sensor inside a Faraday cage or bury a neutrino detector far underground.  However, such vanishing boundary conditions are only physically realizable in local experiments with fields and sources that allow screening, which the gravitational field does not.  In a non-gravitational theory the system $S$ may be insensitive to $C$, but again, this cannot happen in gravity due to the equivalence principle --- all systems interact with the gravitational field.  Finally, in a classical theory one could also in principle just measure all the necessary boundary data.  This is not obviously possible in a gravitational theory that incorporates quantum mechanics, as the amount of information necessary to specify the data at each point on the boundary may violate covariant entropy bounds~\cite{Bousso:2002ju}. 

Given that assumptions about $C$ must exist, in the Alice and Bob information exchange construction necessary for local information objectivity and \v{C}encov's theorem the problem becomes clear: there is no method by which Bob's boundary conditions can be determined from Alice's data since they reside in different causal diamonds. One can give joint boundary conditions on the union of their respective causal boundaries, but this approach extended to all observers rapidly implies an experiment and boundary conditions over the whole spacetime, i.e., at asymptotic infinity.  Alternatively, Alice and Bob can make assumptions about their a priori independent boundaries.  This is what happens in scattering approaches or AdS/CFT, where the asymptotic boundaries are assumed to satisfy symmetries that fix boundary data.  And indeed, in both these approaches information theory and quantum mechanics hold unchanged. For any local experiment however such an argument fails.

In summary, for gravitational theories the act of recording permanent data changes observed systems, and sufficient statistics cannot be generated and passed between observers due to different boundary conditions. Such operations are the information theory equivalent of the fixed stage and bystanding observers of classical, Newtonian physics.  Therefore \v{C}encov's theorem does not apply, the information metric does not have to be the Fisher metric, and the Born rule is not necessarily appropriate for local quantum gravitational experiments~\cite{Berglund:2025qyo}.  Similar arguments can be applied to evade proofs of the Born rule in standard quantum mechanics~\cite{Berglund:2025qyo}. Quantum gravity is not just quantization of classical gravity, but may also entail the generally covariant gravitization of quantum mechanics~\cite{Minic:2003en, Minic:2003nx, Jejjala:2007rn, Freidel:2013zga, Freidel:2014qna, Freidel:2015pka, Berglund:2022skk, Hubsch:2024agh} and of classical information theory itself.  

Phenomenologically, this would imply that state dependent deviations from maximal symmetry may be allowed. The generalized Born rule therefore would contain an expansion of the form (analogous to the expression for the energy density in non-linear optics~\cite{Namdar:2021czo}, where instead of 
$\psi$ one has the electric field)~\cite{Berglund:2023vrm}:
\begin{equation}
  P =    g_{ab}(\psi)\, \psi_a \psi_b \equiv \delta_{ab}\, \psi_a \psi_b + \beta_{abc}\, \psi_a \psi_b \psi_c+\dots
   ,
   \label{e:deformP}
\end{equation}
where $a,b,c$ are state-space 
indices. Geometrically, this generates a Finsler-like modification of the canonical complex projective geometry, $CP^N$ of quantum theory. Distances on $CP^N$ determine the Born rule, and the corresponding geodesic equation is the Schr\"{o}dinger equation. The geodesic equations on the K\"{a}hler--Finsler~\cite{chen2022kahlerfinslermanifoldscurvatures} manifolds would imply consistent evolution equations, compatible with above deformations of the Born rule. The unitary isometries that preserve the Born rule, and lead to conserved probabilities, would be, similarly, consistently modified. 

Modifying the probability rule requires modification of the evolution equations to preserve modified probability measures~\cite{
Helou:2017nsz, Berglund:2023vrm}. Such consistent modified probability and evolution equations already occur in Nambu quantum theory~\cite{Minic:2002pd, Minic:2020zjb, Bhatta:2023biy } and, effectively, in non-linear optics where the background medium is non-trivial and reacts \textit{dynamically} to otherwise linear plane waves~\cite{Namdar:2021czo}. 
To describe the intrinsic triple interference phenomenon 
we 
consider
three distinct 
state-spaces, and also distinguish the corresponding state-functions by a parenthetical superscript~\cite{Berglund:2023vrm}.
 Using the deformation~\eqref{e:deformP} 
 the coupled system of dynamical equations may then be written as (again in analogy with non-linear optics~\cite{Namdar:2021czo})\footnote{In the context of linear optics, and especially, the analysis of polarization, the relevant geometrical structure is the Poincar\'{e} sphere, a direct analog of $CP^{1}$ in quantum theory (the Bloch sphere), which in the context of nonlinear optics also gets deformed into a consistent geometric structure~\cite{Samim_2016}.}:
%
\begin{subequations}
 \label{e:3slits}
\be
 \frac{\rd \psi_a^{(3)}}{\rd \tau}
 =\gamma_{abc}\,\psi_b^{(1)} \psi_c^{(2)},
\label{e:3=12}
\ee
accompanied by (including the canonical linear term)
\be
\frac{\rd \psi_a^{(1)}}{\rd \tau}
 = \delta_{ab}\,\psi_c^{(2)*}
  + \gamma_{abc}\,\psi_b^{(3)} \psi_c^{(2)*},
\label{e:1=23}
\ee
and finally (also including the canonical linear term)
\be
\frac{\rd \psi_a^{(2)}}{\rd \tau}
 = -\delta_{ab}\,\psi_c^{(1)*}
  - \gamma_{abc}\,\psi_b^{(3)} \psi_c^{(1)*}.
\label{e:2=13}
\ee
\end{subequations}
The subscripts $a,b,c$ here label states within the 
distinct superscript-indicated
state-space; 
we proceed assuming that the three state-spaces are isomorphic and so have identically labeled states. (The signs are consistent with real $\psi$'s and the canonical Schr\"{o}dinger equation in the
case of zero $\gamma_{abc}$.)
Corrections as in~\eqref{e:deformP} generate multi-linear interference, which has already been observed~\cite{Namdar:2021czo} in the non-linear optics context and examined in quantum mechanics by Sorkin~\cite{Sorkin:1994dt}.
Much as the canonical quadratic term in~\eqref{e:deformP} defines the standard  probability in quantum mechanics
\be
 P=\d_{ab}\,\j_a\j_b~~\mapsto\quad
 P_2(i,j):=|\j^{(i)}+\j^{(j)}|^2,
\ee
superposing $\j_a\to\j^{(i)}+\j^{(j)}$, the cubical deformation~\eqref{e:deformP} 
leads to, to leading order in $\g$ 
\be
 P_3(1,2,3):=\big|\j^{(1)}+\j^{(2)}+\j^{(3)}\big|^2
 +\Re\big[\g\,(\j^{(1)}+\j^{(2)}+\j^{(3)})^3\big],
\ee
where to lowest order in $\g$ the argument of $\Re$ uses ordinary superposition, uncorrected by further powers of $\g$. 
Thus we have the key measure (properly normalized) of the {\it intrinsic triple interference}, $\widehat\kappa(1,2,3) := \g\,\j^{(1)}\,\j^{(2)}\,\j^{(3)}/
\big( P_2(1,2){+}P_2(1,3){+}P_2(2,3) \big)^{3/2}$~\cite{Berglund:2023vrm}.

One striking observational consequence of this analysis is a modification to the Talbot effect in interferometry~\cite{Berglund:2023vrm}. In the classical Talbot effect, a plane wave diffracts on a grating, generating a grating image at regular distances (the Talbot length) and self images (also called Talbot images) at fractions of the Talbot length~\cite{Berry_2001}. The linear Talbot effect can be observed in the quantum context with matter waves~\cite{Berry_2001}. Numerical simulations~\cite{Berglund:2023vrm} indicate that quantum gravity induced modifications manifest as a significant reduction in the periodic pattern referred to as the Talbot carpet.

Future experiments involving matter-wave diffraction of nanoparticles of masses up to $10^7$ atomic mass units can probe the Talbot effect and search for deviations beyond current limits~\cite{Berglund:2023vrm}.  First, optically trapped nanoparticles can be cooled with parametric laser feedback cooling in a 4K cryostat using an optical tweezer, using parametric feedback cooling techniques already developed by one of us (Geraci). After releasing the particle from the trap and allowing it to expand significantly in free-fall, a light grating pulse with tunable wavelength will be applied.  An imaging system with approximately 10 nm spatial resolution placed below the grating will build up interference patterns one particle at a time based on the location of where the center of mass of the particle lands. The modifications of the predicted interference patterns coupled with the tunability of the grating then will allow parametric searches for intrinsic triple and higher order quantum interference effects~\cite{Berglund:2023vrm}.

Any discovered modification would provide a smoking gun experimental signature
for gravitized quantum theory, in other words, for a dynamical information metric in quantum mechanics coming from quantum gravity. More fundamentally, physics would have taken an additional step towards the incompletely realized dynamical revolution started in both general relativity and quantum mechanics by implementing the gravitational principle of general covariance for both quantum mechanics and information theory~\cite{Minic:2003en, Minic:2003nx, Jejjala:2007rn, Freidel:2013zga, Freidel:2014qna, Freidel:2015pka, Berglund:2022skk, Hubsch:2024agh}.

\paragraph{Acknowledgments:}
We thank N. Bhatta, L. Freidel, P. Geng, J. Kowalski-Glikman, R. Leigh, N. Musoke, A. Smith, and T. Takeuchi 
and we
acknowledge 
the COST Action CA23130.
P.~Berglund thanks the CERN Theory Group for their hospitality and the support of the U.S.\ Department of Energy grant DE-SC0020220.
 T.~H\"ubsch thanks the Mathematics Department of the University of Maryland and the Physics Department of the University of Novi Sad, Serbia, for recurring hospitality and resources. 
 D.~Minic is supported by the Julian Schwinger Foundation and 
the U.S.\ Department of Energy grant
DE-SC0020262, and he thanks Perimeter Institute for hospitality and support. 
A.~Geraci acknowledges support from NSF grants PHY-2409472 and PHY-2111544, DARPA, the John Templeton Foundation, the W.M. Keck Foundation, the Gordon and Betty Moore Foundation Grant GBMF12328, DOI 10.37807/GBMF12328, the Alfred P. Sloan Foundation under Grant No.\ G-2023-21130.

\bibliographystyle{utphys}
\bibliography{Refs}

\end{document}